\documentclass[preprint,aps,tightenlines,floatfix,showpacs]{revtex4}

\usepackage{graphics}
\usepackage{bm}
\usepackage{amsmath}
\usepackage{amssymb}
\begin{document}

\title
{Predicting total and total reaction cross sections using a simple functional 
form}
\author{P. K. Deb}
\email{pdeb@mps.ohio-state.edu}
\affiliation
{Department of Physics, The Ohio State University, Columbus, OH 43210, USA.}
\author{K. Amos}
\email{amos@physics.unimelb.edu.au}
\affiliation
{School of Physics, The University of Melbourne, Victoria 3010, Australia}

\date{\today}
\begin{abstract}
Total reaction cross sections have been predicted for nucleons scattering 
from nuclei ranging in mass 6 to 238 and for projectile energies from just 
above noticeable giant resonance excitation to 300 MeV. 
So also have been the mass variations of those cross sections at selected 
energies when they have been calculated using coordinate space optical 
potentials formed by full folding effective two-nucleon ($NN$) interactions
with one body density matrix elements (OBDME) of the nuclear ground states.
Good comparisons with data result when effective $NN$ interactions defined by 
medium modification of free $NN$ $t$ matrices are used. However, there is a 
simple three parameter functional form that reproduces the partial wave total
reaction cross section values determined from those optical potential 
calculations; a functional form also maps the total scattering cross section
partial wave elements. Adjusting the theoretical defined parameter values has
enabled us to fit the actual measured data values from the scattering involving
(15) nuclei spanning the mass range from $^{9}$Be to $^{238}$U and for proton
energies from 10 to 300 MeV. Likewise total cross sections for neutron cross 
sections for neutron scattering from various nuclei can be equally well 
reproduced. Of import is that the three parameter values vary smoothly with 
mass and energy.
\end{abstract}
\maketitle

\section{Introduction}

Total and total reaction cross sections from the scattering of nucleons by
 nuclei are required in 
a number of fields of study which range over problems in basic science as well
as many of applied nature. Such as in radioactive waste management, in radiation therapy and radiation protection for patients and in many other fields. In 
medical radiotherapy absorbed dose distributions in the patient are needed and 
cannot be measured directly as we should not do nuclear experiment with 
humen body, they must be calculated. For all these purposes an extensive data
bank is necessary. Since we do not have the experimental data for 
all necessary nuclei at all different energies it would be utilitarian if such 
scattering data were
well approximated by a simple convenient function form with which predictions 
could be made for cases of energies and/or masses as yet to be measured.  
It has been shown~\cite{Ma01,Am02,Ken02,De03} that such forms may exist for 
proton total reaction cross sections.  Herein we consider that concept further 
to reproduce the measured total reaction cross sections from protons at  
energies from 10 to 300 MeV and from 
15 different nuclei ranging in mass between $^9$Be and $^{238}$U and total 
cross sections from neutron scattering for 
energies to 600 MeV and from nine nuclei ranging in mass between ${}^6$Li and 
${}^{238}$U. These suffice to show that such forms will also be applicable in 
dealing with other stable nuclei since their neutron total cross sections vary
so similarly with energy~\cite{Ko03}.

Total reaction cross sections for protons and total cross sections 
for neutrons from nuclei have been well 
reproduced by using optical potentials. In particular, the data (to 300 MeV) 
from the same  nuclei we consider, compare quite well with predictions made
using a $g$-folding method to form nonlocal optical potentials~\cite{Amos02}, 
though there are some notable discrepancies. Alternatively, in a recent study 
Koning and Delaroche~\cite{Ko03} gave a detailed specification of 
phenomenological global optical model potentials determined by fits to quite a 
vast amount of data, and in particular to the neutron total scattering cross 
sections we consider herein. 

\section{ Formalism}

The total and total reaction cross sections for nucleons scattering from nuclei can be expressed 
in terms of partial wave scattering ($S$) matrices specified at energies 
$E\propto k^2$, by
\begin{equation}
S^{\pm}_l \equiv S^{\pm}_l(k) = e^{2i\delta^{\pm}_l(k)} =
\eta^{\pm}_l(k)e^{2i\Re\left[ \delta^{\pm}_l(k) \right] }\ ,
\end{equation}
where $\delta^\pm_l(k)$ are the (complex) scattering phase shifts and 
$\eta^{\pm}_l(k)$ are the moduli of the $S$ matrices. The superscript 
designates $j = l\pm 1/2$. In terms of these quantities, the elastic, reaction
(absorption), and total cross sections respectively are given by
\begin{eqnarray}
\sigma_{\text{el}}(E) & = & \frac{\pi}{k^2} \sum^{\infty}_{l = 0} \left\{
\left(l + 1 \right)\left|S^+_l(k) - 1 \right|^2 + l\left|S^-_l(k) - 1\right|^2 
\right\} = \frac{\pi}{k^2} \sum_l \sigma_l^{(el)}\\
\sigma_{\text{R}}(E) & = & \frac{\pi}{k^2} \sum^{\infty}_{l = 0}\left\{ \left(
l + 1 \right) \left[ 1 - \eta^+_l(k)^2 \right] + l \left[ 1 - \eta^-_l(k)^2 
\right] \right\} = \frac{\pi}{k^2} \sum_l \sigma_l^{(R)}\ ,
\label{xxxx}
\end{eqnarray}
and
\begin{eqnarray}
\sigma_{\text{TOT}}(E) & = & \sigma_{\text{el}}(E) + \sigma_{\text{R}}(E)
= \frac{\pi}{k^2} \left[\sigma_l^{(el)} + \sigma_l^{(R)}\right] 
= \frac{2\pi}{k^2} \sum_l \sigma_l^{(TOT)}\ ,
\nonumber\\ 
\sigma_l^{(TOT)} & = &  \left( l + 1 \right) 
\left\{ 1 - \eta^+_l(k)\cos\left( 2\Re\left[ \delta^+_l(k) \right] \right) 
\right\} + l\left\{1 - \eta^-_l(k) \cos\left( 2\Re\left[ \delta^-_l(k) \right] 
\right) \right\}\ .
\label{SumTOT}
\end{eqnarray}
Therein the $\sigma_l^{(X)}$ are defined as partial cross sections of the total
elastic, total reaction, and total scattering itself.   For proton scattering, 
because Coulomb amplitudes diverge at zero degree scattering, only total 
reaction cross sections are measured. Nonetheless study of such 
data~\cite{Am02,De03} established that partial total reaction cross sections 
$\sigma_l^{(R)}(E)$ may be described by the simple function form
\begin{equation}
\sigma_l^{(R)}(E) = (2l+1) \left[1 + e^{\frac{(l-l_0)}{a}}\right]^{-1} + 
\epsilon\ (2l_0 + 1)\ e^{\frac{(l-l_0)}{a}} \left[1 + e^{\frac{(l-l_0)}{a}}
\right]^{-2}\ ,
\label{Fnform}
\end{equation}
with the tabulated values of $l_0(E,A)$, $a(E,A)$, and $\epsilon(E,A)$ all 
varying smoothly with energy and mass.  Those studies were initiated with the
partial reaction cross sections determined by using complex, non-local, 
energy-dependent, optical potentials generated from a $g$-folding 
formalism~\cite{Am00}. While those $g$-folding calculations did not always give
excellent reproduction of the measured data (from $\sim$ 20 to 300 MeV for 
which one may assume that the method of analysis is credible), they did show a 
pattern for the partial reaction cross sections that suggest the simple 
function form given in Eq.~(\ref{Fnform}).  With that form excellent 
reproduction of the proton total reaction cross sections for many targets and
over a wide range of energies were found with parameter values that varied 
smoothly with energy and mass.

Herein we establish that the partial reaction cross sections for scattering of 
protons and total cross sections for scattering of 
neutrons from nuclei can also be so expressed and we suggest forms, at least
first average result forms, for the characteristic energy and mass variations 
of the three parameters involved. Fifteen nuclei, $^9$Be, $^{12}$C, $^{16}$O,
$^{19}$F, $^{27}$Al, $^{40}$Ca, $^{63}$Cu, $^{90}$Zr, $^{118}$Sn, $^{140}$Ce,
$^{159}$Tb, $^{181}$Ta, $^{197}$Au, $^{208}$Pb and $^{238}$U, are considered 
for proton case and nine nuclei, $^6$Li, $^{12}$C, $^{19}$F, 
$^{40}$Ca, $^{89}$Y, $^{184}$W, $^{197}$Au, $^{208}$Pb and $^{238}$U, for
which a large set of experimental data exist, are considered for neutron case.  Also those 
nuclei span essentially the whole range of target mass.  However, to set up an
appropriate simple function form, initial partial total cross sections must be 
defined by some method that is physically reasonable.  Thereafter the measured
total cross-section values themselves can be used to tune details, and of the 
parameter $l_0$ in particular. We chose to use results from $g$-folding optical
potential calculations to give those starting values.

\section{ Results and discussions}

The function form results we display in the following set of figures were 
obtained by starting with $g$-folding model results, for proton case at energies of 10 to 100 MeV in steps of 10 MeV, then to 300 MeV in steps of 50 MeV, and 
for neutron case, at energies of 10 to 100 
MeV in steps of 10 MeV, then to 350 MeV in steps of 25 MeV, and thereafter in
steps of 50 MeV to 600 MeV.  The $g$-matrices used above pion threshold were 
those obtained from an optical potential correction to the BonnB 
force~\cite{Ge98} which, while approximating the effects of resonance terms 
such as virtual excitation of the $\Delta$, may still be somewhat inadequate 
for use in nucleon-nucleus scattering above 300 MeV.  Also relativistic 
effects in scattering, other than simply the use of relativistic kinematics 
in the distorted wave approximation (DWA) approach, are to be expected.  
Nonetheless the DWA results are used only to find a sensible starting set of 
the function form parameters $l_0, a$, and $\epsilon$ from which to find ones 
that reproduce the measured total cross-section data. One must also note that 
the $g$-folding potentials for most of the nuclei considered were formed using 
extremely simple model prescriptions of their ground states. A previous 
study~\cite{Amos02} revealed that with good spectroscopy the $g$-folding 
approach gives much better results in comparison with data than that approach 
did when simple packed shell prescriptions for the structure of targets were 
used. That was also the case when scattering from exotic, so-called nucleon 
halo, nuclei were studied~\cite{Am00,La01}. 

\begin{figure}
\scalebox{0.7}{\includegraphics{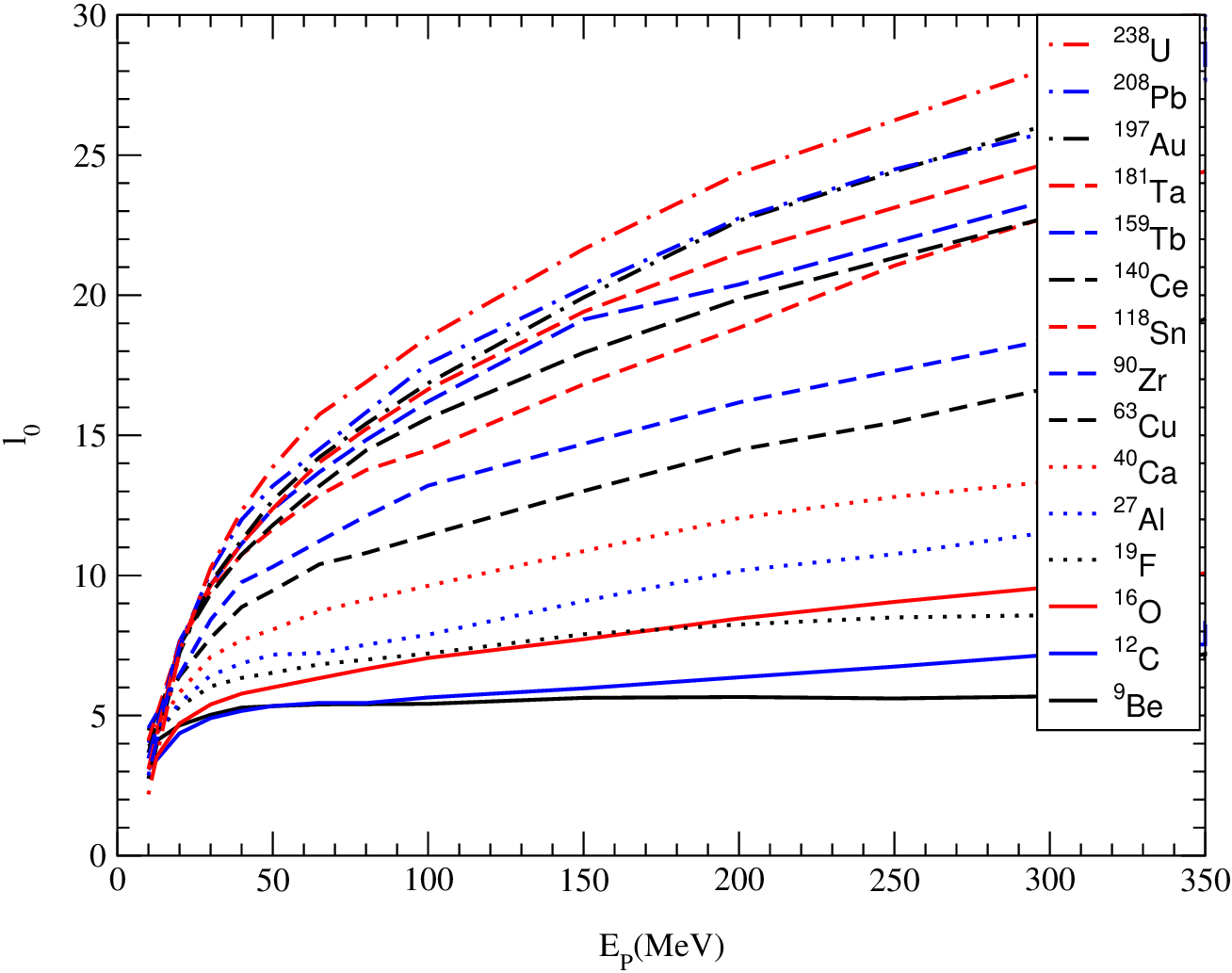}}
\caption{\label{l0-proton-energy}
The parameter values for $l_0$ in the simple functional form for the
$p$-$A$ reaction cross sections.}
\end{figure}

Although on using Eqs.~(\ref{xxxx}) and (\ref{Fnform}) to match values of
(theoretically) calculated total reaction cross sections led~\cite{Am02}
to the three parameters
$l_0(E,A)$, $a(E,A)$, and $\epsilon(E,A)$  having smooth variations
with energy, there are discrepancies between those predictions
and the actual measured data.
Herein we improve the method of selection of those parameter values
to produce more accurate reaction cross section values, while
keeping as smooth a variation with energy of those parameters as
possible.
Specifically, in Eq.~(\ref{Fnform}), we
have set the $\epsilon$ as a constant ($-1.5$) and so independent of energy
and of mass. Further we assume that $a(E,A)$ varies linearly
with the wave vector,
\begin{equation}
k = \frac{1}{\hbar c} \sqrt{E^2 - m^2 c^4}\ ,
\end{equation}
and with the form
\begin{equation}
a(E,A)
\sim 1.02 k - 0.25\ .
\label{fn-values}
\end{equation}
Then $l_0(E,A)$ were adjusted to ensure that all measured total reaction
cross section values are matched by using the function form, Eq.~(\ref{Fnform}).
The resultant optimized values for the
parameter $l_0$ are presented in Fig.~\ref{l0-proton-energy}.
The different patterns and colors of the lines show the $l_0$ values for 
different nuclei. The legend of the graph indicates the nuclei.

\begin{figure}
\centering
\scalebox{0.7}{\includegraphics{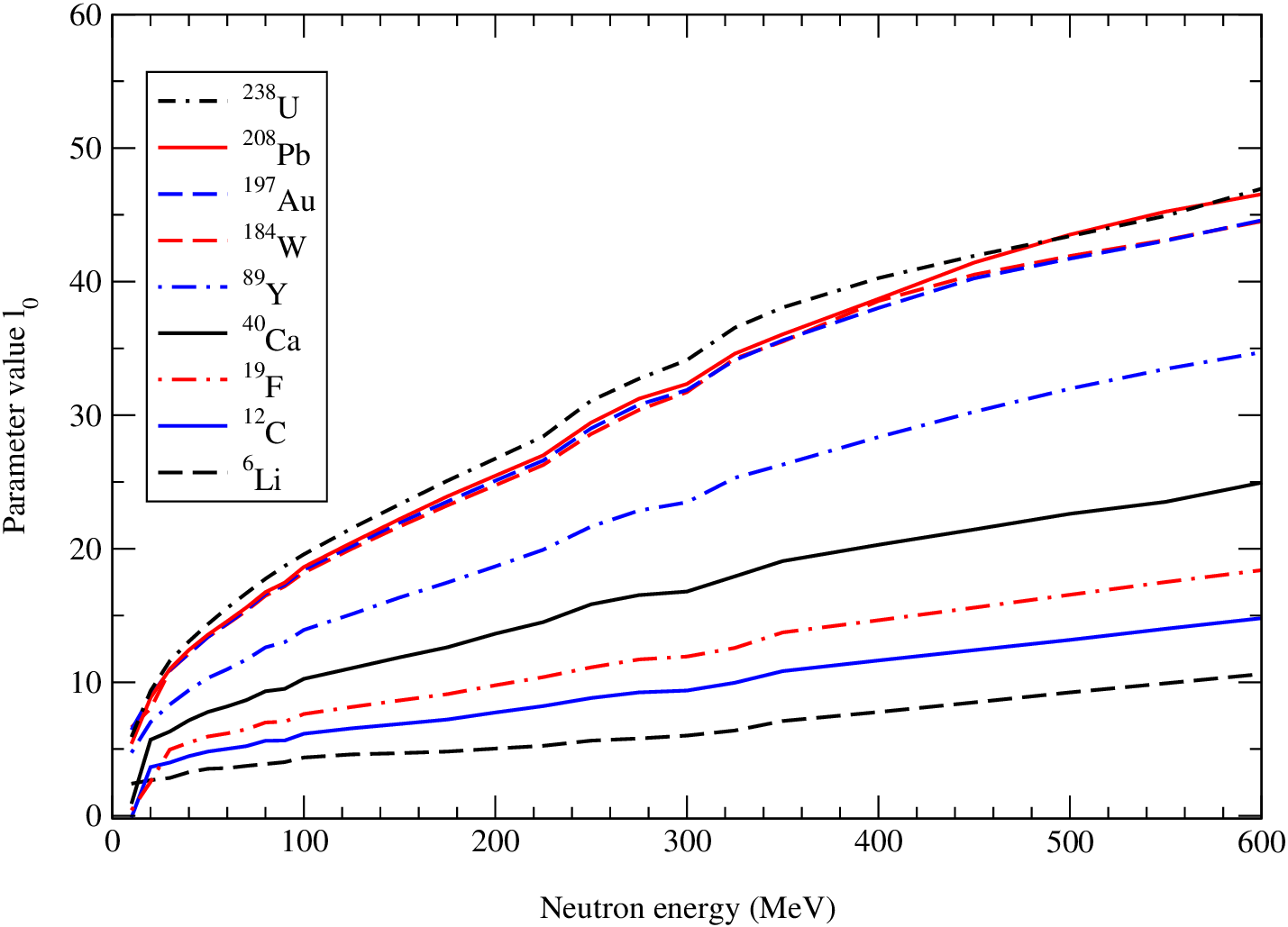}}
\caption{\label{l0vsE} 
The values of $l_0$ that fit neutron total scattering cross-section data from 
the nine nuclei considered and for energies between 10 and 600 MeV. The curves
portray the best fits found by taking a function form for $l_0(E)$.}
\end{figure}

While we have used the partial total cross sections from DWA results for 
neutron scattering from all the nine nuclei chosen and at all of the energies
indicated, 
from the sets of values that result from the fitting process, the two 
parameters $a$ and $\epsilon$ can themselves be expressed by the parabolic 
functions 
\begin{eqnarray}
a\ &=& {\phantom{-}}1.29\ +\ 0.00250\ E\ -\ 1.76\ \times 10^{-6}\ E^2\ , 
\nonumber\\
\epsilon\ &=& -1.47\ -\ 0.00234\ E\ +\ 4.16\ \times 10^{-6}\ E^2\ ,
\label{Eps}
\end{eqnarray}
where the target energy E is in MeV. There was no conclusive evidence for a 
mass variation of them. With $a$ and $\epsilon$ so fixed, we then adjusted the
values of $l_0$ in each case so that actual measured neutron total 
cross-section data were fit using Eq.~(\ref{Fnform}).  Numerical values for
$l_0$ from that process are presented in a table in Ref.~\cite{De04}.
The values of  $l_0$ increase monotonically with both mass and energy and that
is most evident in Fig.~\ref{l0vsE}, where the optimal values $l_0(E)$ are 
presented as different patterned and colored lines. The set for each of the  
masses 
(from 6 to 238) are given by those that increase in value respectively at 600
MeV. While that is obvious for most cases, note that there is some degree of
overlap in the values for ${}^{197}$Au  and for ${}^{208}$Pb.  

\begin{figure}
\scalebox{0.7}{\includegraphics{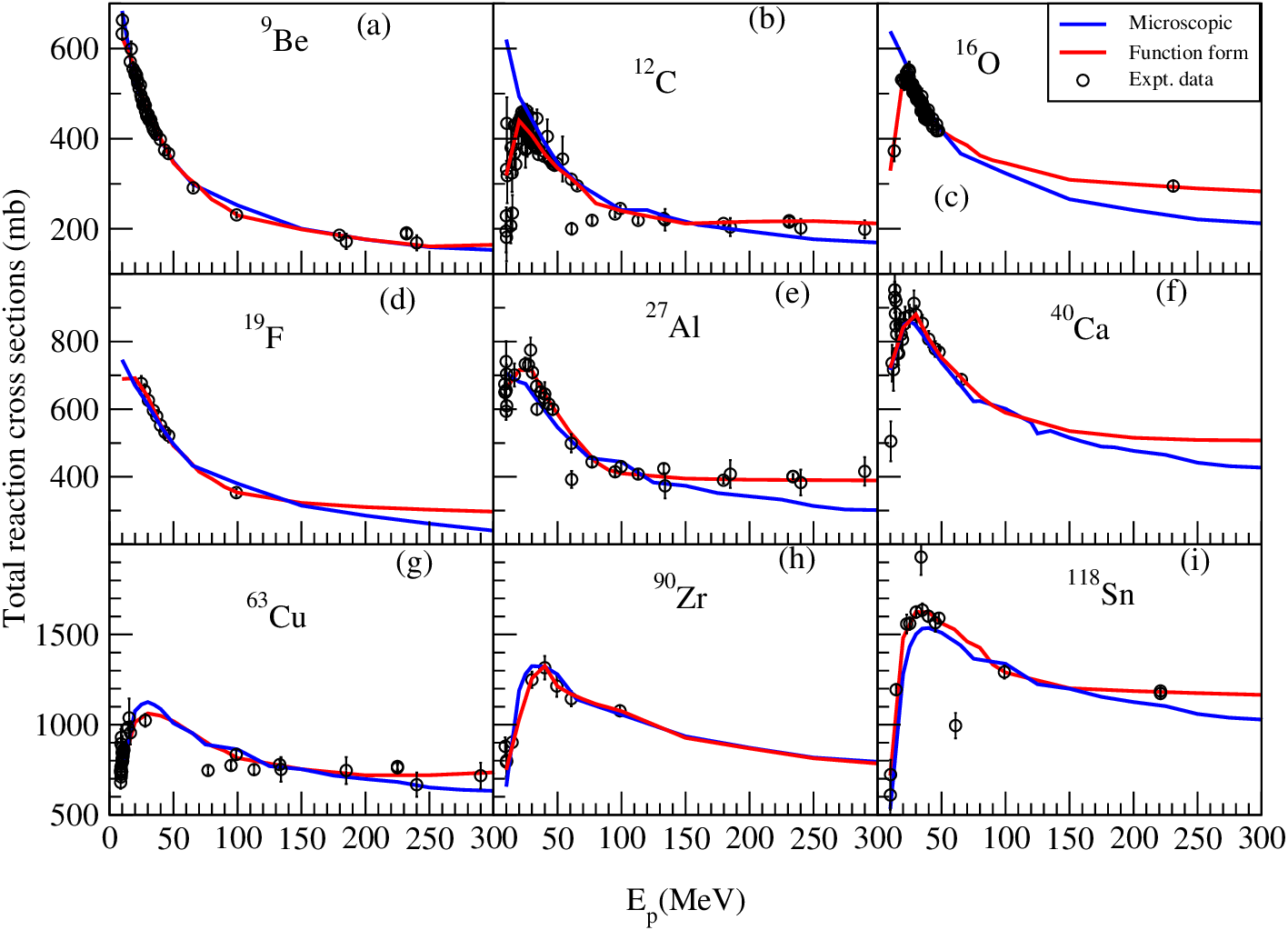}}
\caption{\label{9-118-sigma}
Energy dependence of $\sigma _R$ for proton scattering from
(a) $^9$Be, (b) $^{12}$C, (c) $^{16}$O, (d) $^{19}$F, (e) $^{27}$Al,
(f) $^{40}$Ca, (g) $^{63}$Cu, (h) $^{90}$Zr and (i) $^{118}$Sn.
 Blue lines represent the results obtained from $g$-folding
optical potential calculations while the red curves portray
the values obtained by using simple functional form.}
\end{figure}

\begin{figure}
\scalebox{0.7}{\includegraphics{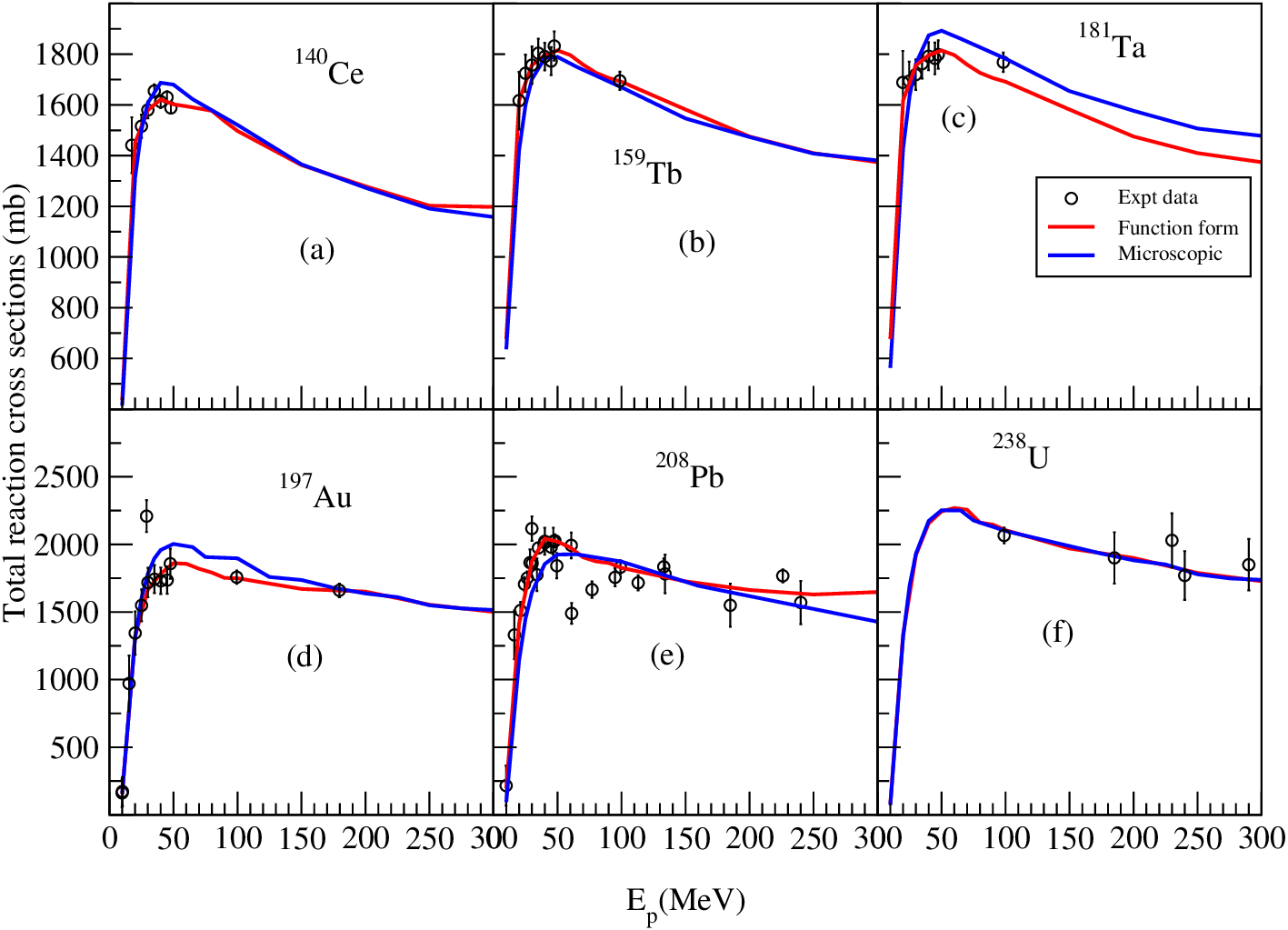}}
\caption{\label{140-238-sigma}
Same as Fig.~\ref{9-118-sigma}, but from
(a) $^{140}$Ce, (b) $^{159}$Tb, (c) $^{181}$Ta, (d) $^{197}$Au, (e) $^{208}$Pb,
 and (f) $^{238}$U.}
\end{figure}

In Figs.~\ref{9-118-sigma} and ~\ref{140-238-sigma},
we compare the total reaction cross sections generated using the simple
functional form and tabled values of $l_0$, and displayed by the red
curves, with those obtained from calculations made using $g$-folding optical
potentials~\cite{Amos02}.  Blue lines represent the predictions obtained from
those microscopic optical model calculations. The experimental
data~\cite{Car96} are depicted by circles.

The results for scattering from $^9$Be, $^{12}$C, $^{16}$O, $^{19}$F, $^{27}$Al,
 $^{40}$Ca, $^{63}$Cu, $^{90}$Zr and $^{118}$Sn are displayed in segments 
(a) through (i)
of Fig.~\ref{9-118-sigma} respectively.
In segment (a), the data are well reproduced by both the $g$-folding
predictions resulting from the folding with the $^9$Be ground state OBDME
found with (0+2)$\hbar\omega$ spectroscopy and by those obtained with the
simple functional form method.

Calculated p-$^{12}$C reaction cross sections are compared with the
experimental data in segment (b) of Fig.~\ref{9-118-sigma}. The reaction cross
sections obtained from $g$-folding calculations are in good
agreement with the experimental data but only in the energy range above 20 MeV.
On the other hand, the results obtained from the simple functional
method are excellent for all energies, replicating the data
very well even in the lower energy range from 10 MeV. There are
two data points, at 61 MeV and at 77 MeV, in disagreement with the
calculated results however. But, as noted previously~\cite{De01},
these data points should be discounted.

Predictions for p-$^{16}$O and for p-$^{19}$F scattering are compared
with the data in segments (c) and (d) of Fig.~\ref{9-118-sigma}.
For p-$^{16}$O case, there are many
data points at the energies between 20 to 40 MeV. Predictions from $g$-folding
calculations while replicating the data very well at and above 25 MeV,
overestimate at lower energies. That $g$-folding result also underestimates
the datum at 250 MeV; the sole datum above 50 MeV. In contrast, the results
obtained from
the simple functional method  are in excellent agreement with the experimental
data at all energies. For p-$^{19}$F, although $g$-folding calculations
reproduce the data very well, the simple functional form method gives slightly
more accurate predictions.

Total reaction cross section
predictions for p-$^{27}$Al and for p-$^{40}$Ca
are compared with the experimental data in segments (e) and (f) of
Fig.~\ref{9-118-sigma}. Again while  $g$-folding calculations reproduce the data
quite well to 200 MeV, three data points at 180 to 300 MeV are not matched.
The $g$-folding results underestimate them noticeably. But predictions from
simple functional form replicate the data very well at all
energies. One data point at 61 MeV is exceptional in the set.
With $^{40}$Ca, the folding model approach is not expected to be reliable
at the energies in the range 10 to 20 MeV, as is the case with $^{12}$C,
since for excitation energies of that range, both nuclei have
clearly discrete spectra.  That is true for most light mass nuclei but little
or no total reaction cross sections have been reported for them.  Indeed the
reaction data from both $^{12}$C and from $^{40}$Ca show rather sharp
resonance-like features below 20 MeV. Both the $g$-folding calculations and
functional form calculations reproduce the rest of the $^{40}$Ca data very well.

For  $^{63}$Cu, our predictions at low energies may be slightly
too small and the parameter sets driven too severely by the sole datum
at 30 MeV in the range 20 to 70 MeV.  Also the data in the range 100
to 300 MeV are quite scattered but the simple functional form gives
a good average result.

In segment (h) of Fig.~\ref{9-118-sigma}, the predicted total reaction cross
sections from p-$^{90}$Zr scattering are compared with the experimental data.
Results from $g$-folding calculations are in very good agreement with  the
data although the data value at 30 MeV is overestimated.
The results obtained from
the simple functional form are in excellent agreement with the experimental
data at the few energies measured, but the shape is not optimally smooth.
 Lack of data meant that we had to use the $g$-folding
values to specify the functional form.  That is also the case with masses
140, 159, and 181.

The p-$^{118}$Sn total reaction cross section results are given in segment (i)
of Fig.~\ref{9-118-sigma}, where the two predictions again
are compared with the data.
Although not as good as the results found for scattering from light mass
nuclei, the $g$-folding potential still gives reasonable shape prediction.
However, the model underestimates the data by 5 to 10\%. But predictions from
the simple functional
form  model form an excellent reproduction of the data at all energies except
61 MeV. This 61 MeV data point is again exceptional being much smaller
than other data and the predictions as in the cases of $^{12}$C and
$^{27}$Al.

Predictions for p-$^{140}$Ce, p-$^{159}$Tb, p-$^{181}$Ta, p-$^{197}$,
p-$^{208}$ and for p-$^{238}$U scattering
are compared with the (limited amount of) data in segments (a), (b), (c), (d), 
(e) and (f) of
Fig.~\ref{140-238-sigma} respectively. The $g$-folding calculations give
very good agreement with that data for p-$^{140}$Ce, slightly underestimate
the data for p-$^{159}$Tb, and for p-$^{181}$Ta,
underestimate data at the energies to 20 MeV and overestimate data in the
energy range 40 to 60 MeV. In all cases results predicted by the simple
functional form are  excellent reproductions of the experimental data.

For $^{197}$Au, the $g$-folding optical potential calculations
are in good agreement with most data; the 29 MeV datum grossly
underestimated by the calculations. But that data point is also at odds with
the energy trend of the other data. Save for that 29 MeV value,
the simple functional form gives even better predictions of the data.
The energy variation of the p-$^{208}$Pb reaction cross sections is shown
in segment (e) of Fig.~\ref{140-238-sigma} where the predictions
from $g$-folding optical potential calculations and from the
simple functional form
calculations are compared with a fairly  extensive set of experimental data.
In making the $g$-folding potentials,
we have used Skyrme-Hartree-Fock  wave functions~\cite{Br00} which
have been shown to be more
realistic~\cite{Amos02} than  simple oscillator model ones.
Still such  $g$-folding calculations
underestimate the data up to 50 MeV. But simple functional form calculations
are in excellent agreement with the experimental data, save that
data values at 30, 61 and 77 MeV again are exceptional.
The predictions of total reaction cross sections for p-$^{238}$U scattering
from the $g$-folding optical potential calculations and from using the simple
functional form are compared with the few data
in segment (f) of Fig.~\ref{140-238-sigma}.  Given the lack of data
the two results are virtually identical.

As a final note regarding many of the exceptional data values so defined
in the foregoing, Menet {\it et al.}~\cite{Men71} argue  that there may be
a systematic error in the studies reported in those experiments.

\begin{figure}
\centering \scalebox{0.7}{\includegraphics{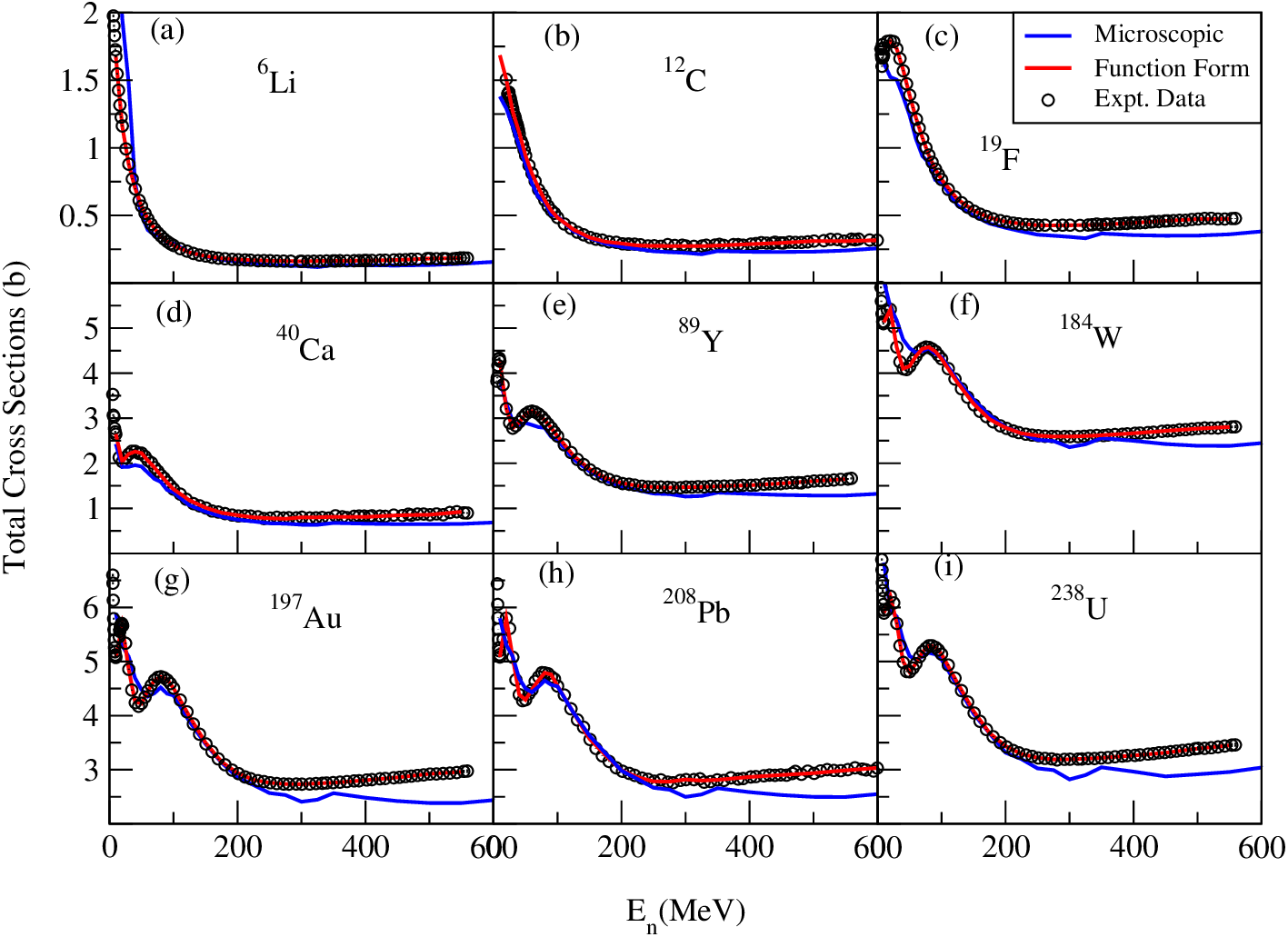}}
\caption{\label{nTotX}
Total cross sections for neutrons scattered from (a) $^6$Li, (b) $^{12}$C, 
(c) $^{19}$F, (d) $^{40}$Ca, (e) $^{89}$Y, (f) $^{184}$W, 
(g) $^{197}$Au, (h) $^{208}$Pb and (i) $^{238}$U.}
\end{figure}

The total neutron scattering cross sections generated using the function form
for partial total cross sections with the tabled values of $l_0$
and the energy function forms of Eq.~(\ref{Eps}) for $a$ and $\epsilon$,
are shown in Fig.~\ref{nTotX}.  
They are displayed by the red lines that closely match the data which 
are portrayed by  circles. The data that was taken from a survey by 
Abfalterer~{\it et al.}~\cite{Abf01} which includes data measured at LANSCE 
that are supplementary and additional to those published earlier by 
Finlay~{\it et al.}~\cite{Fin93}. For comparison we show results obtained from 
calculations made using $g$-folding optical potentials~\cite{Amos02}. Blue 
lines represent the predictions obtained from those microscopic optical 
potential calculations. Clearly for energies 300 MeV and higher, those 
predictions fail.
 
The total cross sections for neutrons scattered from the  nuclei
considered are compared with data in Fig.~\ref{nTotX}.
Again 
the $g$-folding potential results are displayed by the blue curves while 
those of the function form are shown by the red curves.
We show in segment (h) of  Fig.~\ref{nTotX}, the results for neutron scattering 
from ${}^{208}$Pb. In this case we used Skyrme-Hartree-Fock model (SKM*) 
densities~\cite{Br00} to form the $g$-folding optical potentials. That 
structure when used to analyze proton and neutron scattering differential 
cross sections at 65 and 200 MeV gave quite excellent results~\cite{Ka02}.  
Indeed those analyzes were able to show selectivity for that SKM* model of 
structure and for the neutron skin thickness of 0.17 fm that it proposed. 
Using the SKM* model structure, the $g$-folding optical potentials gave the 
total cross sections shown by the blue curve in segment (h) of 
Fig.~\ref{nTotX}. Of 
all the results, we believe these for ${}^{208}$Pb point most strongly to a 
need to improve on the $g$-folding prescription as is used currently when 
energies are at and above pion threshold.  Nonetheless, it does do quite well
for lower energies, most notably giving a reasonable account of the 
Ramsauer resonances~\cite{Ko03} below 100 MeV. However, as with the other 
results, these $g$-folding values serve only to define a set of partial 
cross sections from which an initial guess at the parameter values of the 
function form is specified. With adjustment that form produces the red curve
shown in segment (h) of  Fig.~\ref{nTotX}, which is an excellent reproduction of 
the data,
as it was designed to do.  But the key feature is that the 
optimal fit parameter values still vary smoothly with mass and energy.

Without seeking further functional properties of the parameters, one could 
proceed as we have done this far but by using many more cases of target mass
and scattering energies so that a parameter tabulation as a data base may be 
formed with which any required value of total scattering cross section might 
be reasonably predicted (i.e. to within a few percent) by suitable 
interpolation on the data base, and the result used in Eq.~(\ref{Fnform}).
\section{Conclusions}
Measured reaction cross sections for 10 to 300 MeV proton scattering from 
nuclei ranging in mass from $^9$Be to $^{238}$U are well reproduced by 
calculations made using the g-folding model of the optical potential. Measured
total cross sections for 10 to 600 MeV neutron scattering from nuclei ranging
in mass from $^6$Li to $^{238}$U are well reproduced. Those calculated results
gave a set of partial reaction cross sections in each case that vary smoothly
with target mass and energy.
Those variations are well reproduced by a simple three parameter function. 
Using the simple function form total and total reaction cross sections are 
well reproduced. This simple functional method can be used to estimate cross 
sections for many useful applications, such as, medical radiotherapy.

\begin{acknowledgments}
This research was supported by a research grant from the Australian Research 
Council and also by the National Science Foundation under Grant No. 0098645.
\end{acknowledgments}

\bibliography{Aps-april04}

\end{document}